\documentclass[12pt]{iopart}
\usepackage{iopams}  
\usepackage{graphicx,url}
\usepackage{astroforiop}

\newcommand{\revised}[2]{#2}

\begin{document}

\article[The supermassive black hole in the Galactic Center]{\review{}}{Towards the event horizon - the supermassive black
  hole in the Galactic Center}

\author{H Falcke$^{1,2,3}$ and S B Markoff$^4$}

\address{$^1$ Department of
  Astrophysics, Institute for Mathematics, Astrophysics and Particle
  Physics (IMAPP), Radboud University Nijmegen,
  P.O. Box 9010, 6500 GL Nijmegen, The Netherlands
}

\address{$^2$ ASTRON, Oude
  Hoogeveensedijk 4, 7991 PD Dwingeloo, The Netherlands
} 

\address{$^3$ Max-Planck-Institut f\"ur Radioastronomie, Auf dem H\"ugel 69, 53121
Bonn, Germany}

\address{
$^4$ Astronomical Institute 'Anton Pannekoek', University of Amsterdam, Postbus 94249, 1090 GE Amsterdam, The Netherlands
}

\ead{\mailto{h.falcke@astro.ru.nl}, \mailto{s.b.markoff@uva.nl}}

\begin{abstract}
  The center of our Galaxy hosts the best constrained supermassive
  black hole in the universe, Sagittarius A* (Sgr~A*). Its mass and
  distance have been accurately determined from stellar orbits and
  proper motion studies, respectively, and its high-frequency radio,
  and highly variable near-infrared and X-ray emission originate from
  within a few Schwarzschild radii of the event horizon.  The theory
  of general relativity (GR) predicts the appearance of a black hole
  shadow, which is a lensed image of the event horizon. This shadow
  can be resolved by very long baseline radio interferometry and test
  basic predictions of GR and alternatives thereof. In this paper we
  review our current understanding of the physical properties of
  Sgr~A*, with a particular emphasis on the radio properties, the
  black hole shadow, and models for the emission and appearance of the
  source. We argue that the Galactic Center holds enormous potential
  for experimental tests of black hole accretion and theories of
  gravitation in their strong limits.
\end{abstract}

\submitto{\CQG}

\maketitle

\section{Black holes in astrophysics}
The defining feature which distinguishes black holes from any other
physical object is the event horizon. This “one-way” membrane in the
fabric of spacetime defines not only the boundary between regions that
are causally disconnected, but it is also the border where time and
space exchange their nature. However, can one experimentally
demonstrate that an event horizon exists?

Black hole candidates come in two fundamental mass classes: stellar
black holes, which are essentially the collapsed cores of massive
stars that either first exploded as supernovae or directly collapsed,
and supermassive black holes (SMBHs), which reside in the nuclei of
most, if not all, galaxies. If $M_{\rm bh}$ is the mass of the black
hole, its characteristic scale is set by the size of the event horizon
in the non-spinning case, the Schwarzschild radius $R_{\rm S}=2 G
M_{\rm BH} /c^2\sim 3$ km $(M_{\rm bh}/M_\odot)$, where $G$ is the
gravitational constant, $c$ is the speed of light, and
$M_\odot=2\times10^{30}$ kg is the mass of the sun.  Stellar-mass
black holes have masses of order $10 M_\odot$ and thus sizes of
several tens of kilometers. They are typically found at distances of a
few kpc (1 kpc = 1000 pc = $3\times10^{16}$ km). \revised{17}{There
  could be up to $10^8$ stellar BHs in our Galaxy
  \cite{BoissierPrantzos1999a,Samland1998a}, of which only some $10^5$
  would be in binaries using estimates --- based on typical duty cycles
  and the few hundred that have actually been detected
  \cite{Jonkeretal2011}}.

Supermassive black holes have masses between $\sim10^6-10^{10}
M_\odot$ and can be detected in galaxies from Mpc to Gpc distances.
The angular size of $R_{\rm S}$ for a black hole at distance $D$ is
$\theta_{\rm RS}=0.1\;{\rm nanoarcsec}\, (M_{\rm bh} /10M_\odot) ({\rm
  kpc}/D)$. For stellar mass black holes, $R_{\rm S}$ lies well below
the resolution capabilities of any current technology.  Although
SMBHs are intrinsically much bigger they are also much further away,
which makes their angular size generally also too small to be resolved
by any telescope.

Hence, it is worthwhile to turn to the center of our own Galaxy, which
hosts the closest candidate SMBH. Here, telescopes are
  able to resolve the gravitational sphere of influence\footnote{\revised{11}{The
  region where the gravitational potential of the BH dominates over
  the potential due to stars.}} and the outer
  accretion region of the central source so that many basic parameters
  can be constrained --- some to the second digit, others within an
order of magnitude, but always with a level of confidence that is
often impossible to achieve for more distant objects. It is here where
we can hope to develop an ideal ``Laboratory for Magnetohydrodynamics
and General Relativity'' \cite{FalckeMarkoffBower2011a}.

\section{The black hole in the Galactic Center: observational properties}

Based on observations of other active SMBHs (also called Active
Galactic Nuclei; AGN), Lynden-Bell \& Rees \cite{Lynden-BellRees1971}
 already proposed in 1971 to look for a compact radio source in the
center of our own Milky Way, which was then indeed discovered three
years later by the NRAO interferometer at Green Bank
\cite{BalickBrown1974} and then by Westerbork
\cite{EkersGossSchwarz1975}. This ``compact radio source in the
Galactic Center'' became later known as ``Sagittarius A*'' (Sgr A*)
and by now is the best constrained SMBH candidate we know of (see
\cite{GenzelTownes1987,MeliaFalcke2001a,GenzelEisenhauerGillessen2010a,MorrisMeyerGhez2012a}
for reviews and \cite{FalckeHehl2002a} for a textbook).

What makes Sgr~A* so special is its proximity at only 8 kpc, coupled
with its large mass of about 4 million solar masses, yielding
$\theta_{\rm RS}=10\;\mu{\rm as}$. \revised{18}{In fact, Sgr~A*'s
  visible event horizon, affected by lensing in it own gravitational
  potential, is predicted to be} $\sim5\theta_{\rm
  RS}\simeq50\;\mu{\rm as}$ \cite{FalckeMeliaAgol2000}, subtending the
largest angle on the sky compared to any other black hole in the
universe \cite{JohannsenPsaltisGillessen2012a}.

\subsection{Mass of Sgr~A*}
Particularly unique is how accurately the mass of Sgr~A* has been
determined.  Since optical radiation from the Galactic Center is
completely absorbed\footnote{Meaning that the true location of the
  Galactic Center was unknown until it was detected by radio telescopes
  in the 1950s, which then led to a significant revision of the
  Galactic coordinate system by $>30^\circ$ in latitude
  \cite{BlaauwGumPawsey1960a}. Still today the official origin of the
  Galactic Center is slightly offset from the black hole.}, \revised{12}{the only
observing bands where Sgr A* is clearly detected are radio (including
sub-mm waves), the near- and mid- infrared (NIR/MIR), and X-rays}.

\begin{figure}
\includegraphics[width=0.49\textwidth]{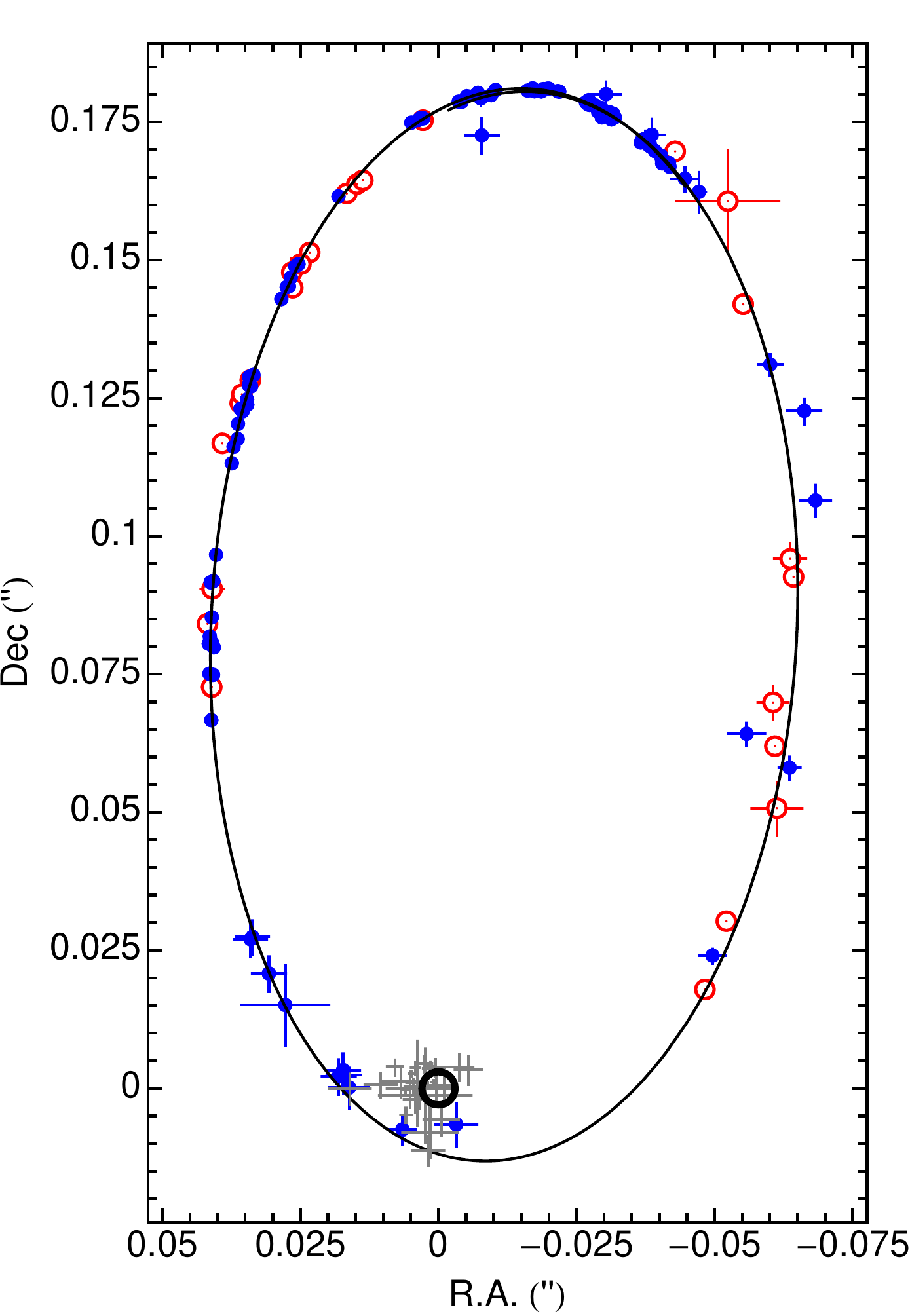}\includegraphics[width=0.475\textwidth]{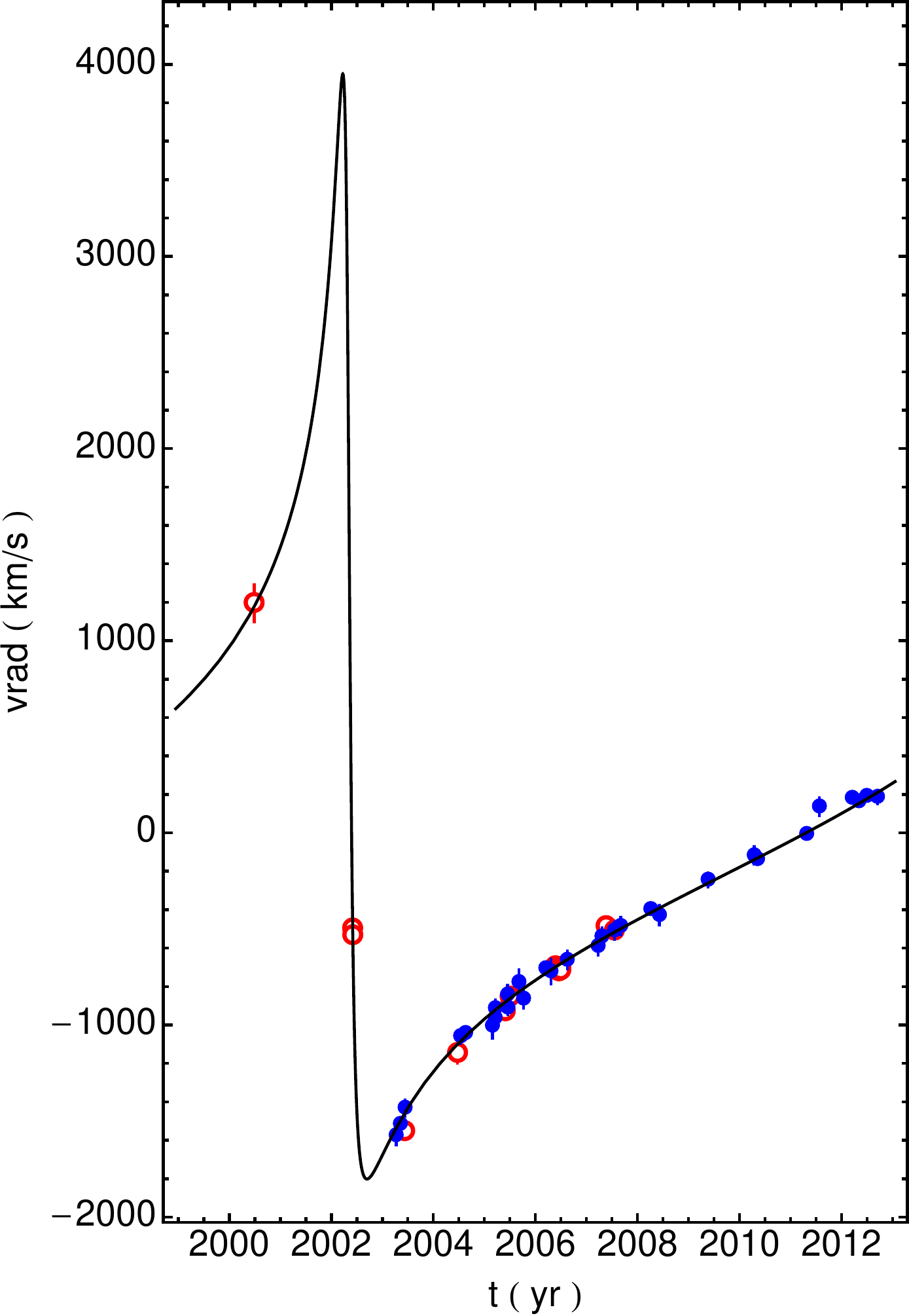}
\caption{\label{fig:orbits} Measured locations and fitted orbit of the
  star S2 around Sgr~A* (left) and its radial velocity (right), taken
  from Genzel, Eisenhauer, \& Gillessen
  \cite{GenzelEisenhauerGillessen2010a}. \revised{8}{The radio position of Sgr~A*
  is marked by a black circle and NIR flares from Sgr A* by grey
  crosses. Blue, filled circles denote the NTT/VLT points of Gillessen
  et al. (2009a,b, updated to 2010), and open and filled red circles
  are the Keck data of Ghez et al. (2008) corrected for the difference
  in coordinate system definition (Gillessen et al. 2009a). } }

\end{figure}

The first evidence for a central dark mass of a few million solar
masses came from the MIR (e.g.,
\cite{LacyTownesGeballe1980,SerabynLacyTownes1988a}) and radio
recombination line observations of gas streamers around Sgr~A*
\cite{RobertsGoss1993a,ZhaoBlundellMoran2010a}.  The final
breakthrough, however, came with the development of adaptive optics
for ground-based facilities. This provided very high spatial
resolution and the ability to detect individual stars in orbit around
Sgr~A* \cite{EckartGenzelHofmann1993a}.  After it was demonstrated
that proper motions of stars could be detected
\cite{EckartGenzel1996,GhezKleinMorris1998}, various groups have been
following the motions of individual stars over several decades and
have used them to map out the gravitational potential within which
they move (e.g.,
\cite{GhezMorrisBecklin2000,Genzeletal2000,GhezSalimWeinberg2008,GillessenEisenhauerTrippe2009a}).
These stars show textbook-like Keplerian elliptical orbits (Figure
\ref{fig:orbits}), with orbital speeds reaching up to 10,000 km
$s^{-1}$ and orbital periods as short as 12-16 years
\cite{SchodelOttGenzel2002,GillessenEisenhauerFritz2009,MeyerGhezSchodel2012a}.

These studies show that indeed the gravitational potential in the central
parsec of the Milky Way must be dominated by a single point source of
mass $M_{\rm BH}=4.3(\pm0.4)\times10^6 M_\odot$
\cite{GhezSalimWeinberg2008,GillessenEisenhauerTrippe2009a}, which is
concentrated within a few hundred $R_{\rm S}$ of Sgr~A*. The limiting
factor is the precise localization of Sgr~A* within the NIR coordinate
frame \cite{MentenReidEckart1997a} and is only good to within $\sim2$ mas
(see Table~4 in \cite{GillessenEisenhauerTrippe2009a}; mas =
milliarcsecond), i.e., $\sim 200\, R_{\rm s}$.

Radial velocities for many stars can be determined from
high-resolution spectroscopy. Combined with proper motions of the same
stars this yields a geometric distance measure of $D=8.3(\pm0.4)$~kpc
\cite{EisenhauerSchoedelGenzel2003,GhezSalimWeinberg2008,GillessenEisenhauerTrippe2009a}.
\revised{7}{The longterm astrometric accuracy of these observations
  can be as good as 0.15-0.3 mas = 15-30 $R_{\rm s}$}. Nonetheless,
the distance uncertainty still provides a significant source of
systematic uncertainty for the mass.  \revised{}{Most recent VLBA
  observations of many masers in the Galaxy yield a Galactic Center
  distance of $D=8.35(\pm0.15)$~kpc \cite{ReidMentenBrunthaler2013a},
  consistent with the position of Sgr~A* and providing smaller error
  bars.}

The final piece of evidence for the association of a dark mass with
Sgr~A* comes from its own proper motion. Very long baseline
interferometry (VLBI) using radio telescopes has been able to track
the position of the source over several years to within a fraction of
a mas per epoch
\cite{ReidReadheadVermeulen1999,ReidBrunthaler2004a}. Sgr~A*'s motion
on the sky is entirely consistent with the solar system motion around
the center of the Galaxy, i.e., the radio source is indeed in the
Galactic Center and not in the fore- or background, and any motion
perpendicular to the projected motion is $<0.4(\pm
0.9)$~km/sec. Compared to the high velocities of stars, this implies
that more than 10\%, if not all, of the dark mass is indeed associated
with the radio source.

\subsection{Spectrum of Sgr~A*}
The apparent size and emission spectrum of Sgr~A* are intimately
linked and are crucial for understanding the source as a whole.

\begin{figure}
\begin{center}
\includegraphics[width=0.65\textwidth]{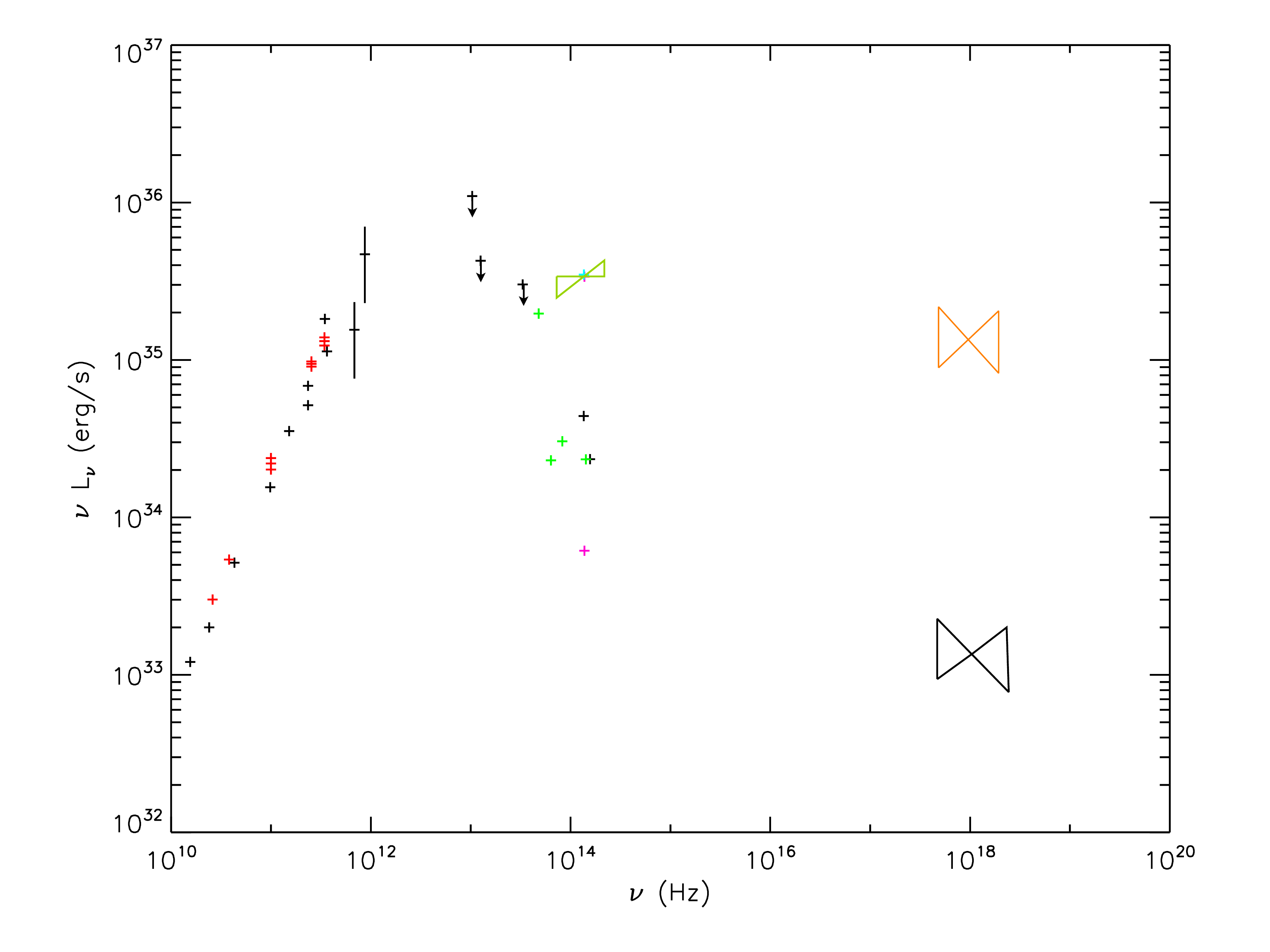}
\end{center}
\caption{\label{fig:spectrum} \revised{16}{Broad-band spectrum of Sgr
    A* compiled by \cite{Dibietal2013}, showing both the quiescent and
    flaring states.  The black data points represent an average
    spectrum, compiled from \cite{FalckeGossMatsuo1998},
    \cite{ZhaoYoungHerrnstein2003}, with NIR upper limits from
    \cite{SerabynCarlstromLay1997} and \cite{Hornsteinetal2002}.  The
    black X-ray “bowtie” is dominated by the outer accretion flow on
    scales of the Bondi radius \cite{BaganoffMaedaMorris2003a}, likely
    containing a $\sim 10\%$ contribution to this flux from Sgr A*
    itself \cite{Neilsenetal2013}.  The red radio data are taken from
    ALMA (Brinkerink \& Falcke, in prep.).  The green IR data points
    are from \cite{Schoedeletal2011}, and the pink and cyan
    representative IR lower/upper limits are taken from
    \cite{Ghezetal2004}, \cite{GenzelOttEckart2003}, and
    \cite{Dodds-Edenetal2011}. The green bowtie is from
    \cite{Bremeretal2011}, and is one of the few IR detections with a
    slope determination that is simultaneous with an X-ray detection.  The
    orange bowtie is from the largest X-ray flare detected so far with \textit{Chandra}
    \cite{NowakNeilsenMarkoff2012a}, and the (dark/light) blue data
    points are two flares observed with NuSTAR
    \cite{Barriereetal2013}. NuStar data is not available in the arXiv version.}}
\end{figure}

Combining all radio data on Sgr~A* one finds that the radio flux
density $S_\nu$, shows a flat-to-inverted spectrum, i.e., it rises
slowly with frequency \revised{16}{with the power peaking around
  $10^{12}$ Hz in the submm band.  Figure~\ref{fig:spectrum} shows the
  broad-band spectrum of Sgr A* as $\nu\times L_\nu$, where $L_\nu= 4
  \pi D_{\rm Sgr~A*}^2 S_\nu$. This frequently used representation
  shows the power flux, and means that the flat radio spectrum becomes
  steeply rising in such a plot.}. At GHz frequencies one has
$S_\nu\propto\nu^\alpha$ and $\alpha\sim0.3\pm0.1$ (which becomes
$\sim1.3$ in a $\nu L_\nu$ plot). The spectrum continues towards low
frequencies ($\sim$300 MHz) with no sign of absorption
\cite{RoyPramesh-Rao2004a,NordLazioKassim2004a}.

At higher frequencies the spectrum extends into the sub-THz
(i.e., sub-mm wavelength) regime
\cite{ZylkaMezger1988,MezgerZylkaSalter1989,SerabynCarlstromLay1997,FalckeGossMatsuo1998},
where the spectrum peaks and then suddenly cuts-off. The so-called ``submm
bump'' is due to synchrotron emission in transition from being
optically thick to thin, and simple arguments demonstrate that this
emission can only arise from the most compact regions several $R_{\rm
  S}$ in diameter \cite{FalckeGossMatsuo1998}. For self-absorbed
synchrotron sources, higher frequencies are typically emitted at
smaller scales, and suspiciously the high-frequency peak in the spectrum of Sgr~A*
implies a scale on the order of the event horizon 
\cite{FalckeBiermann1994a}. Hence, it was argued \cite{FalckeGossMatsuo1998} that the event horizon of
Sgr~A* could be imaged against the ``background'' of this emission,
using VLBI at sub-mm wavelengths (``submm-VLBI'').

\subsection{Size and structure of Sgr~A*}
\begin{figure}
\begin{center}
 \includegraphics[width=0.49\textwidth]{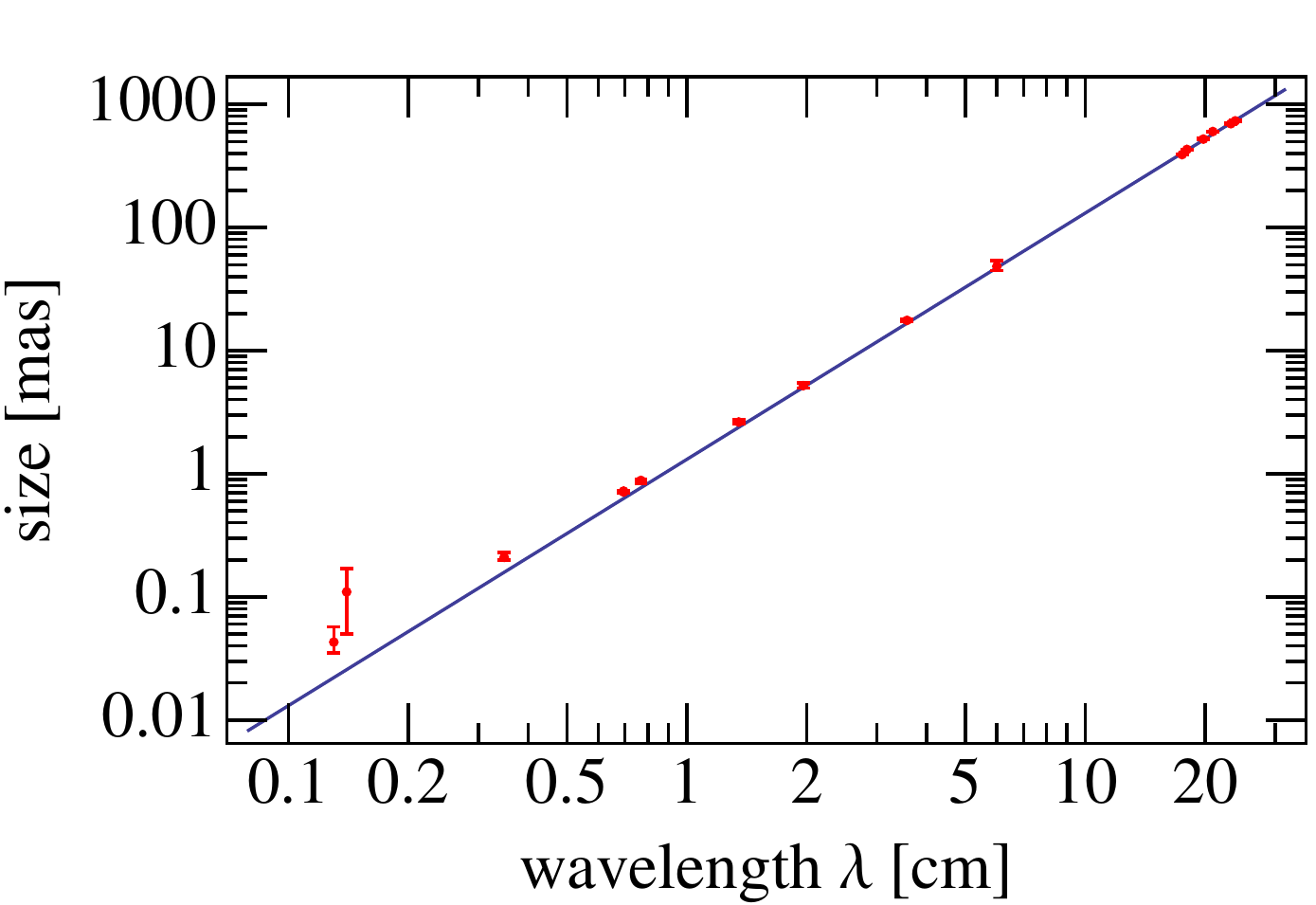} \hfill \includegraphics[width=0.49\textwidth]{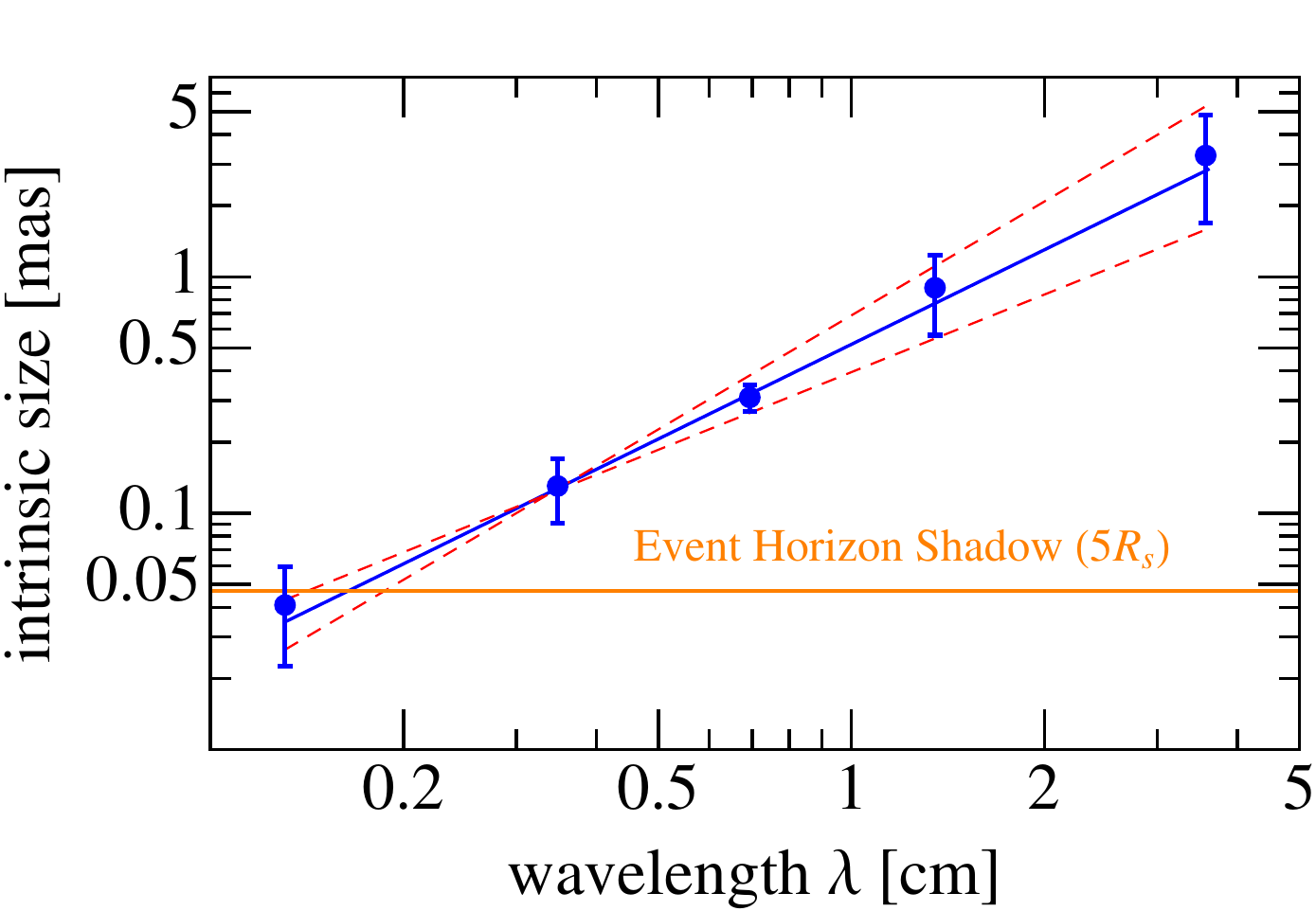} 
 \caption{\label{fig:size} Left: measured major axis size of Sgr~A* as
   function of wavelength measured by various VLBI experiments, the solid
   line indicates the $\lambda^2$ scattering law. Right: Derived
   intrinsic size of Sgr~A* after subtraction of the
   scattering law using all available VLBI data (see
   \cite{FalckeMarkoffBower2009a} for details). The dashed line
   indicates the systematic uncertainties due to different
   normalizations of the scattering law, \revised{18}{the predicted event horizon size due to
   lensing effects is indicated \cite{FalckeMeliaAgol2000}}. Figure taken from
   \cite{FalckeMarkoffBower2011a}.}
   \label{figsgrsize}
\end{center}
\end{figure}

One of the first ``disappointments'' of interferometric measurements
of Sgr~A* was that the measured source size and structure were not
intrinsic to the source itself. The shape of Sgr~A* is blurred into an
east-west oriented ellipse of axial ratio 2:1 caused by scattering of
the radio waves by electrons in the interstellar medium between us
and the Galactic Center
\cite{DaviesWalshBooth1976a,LoBackerEkers1985a,LoShenZhao1998,BowerGossFalcke2006}. \revised{-}{Interestingly,
  recent data from a similarly scattered pulsar near Sgr A*
  \cite{EatoughFalckeKaruppusamy2013a,SpitlerLeeEatough2013a} 
  suggest that the location of this scattering screen is actually not
  directly in the Galactic center \cite{BowerDellerDemorest2013a}}.
The observed size of Sgr~A* follows a $\lambda^2$ law, such that the
scatter-broadened angular size is (\cite{FalckeMarkoffBower2009a}, see
also Figure~\ref{fig:size}, left)

\begin{equation}
\phi_{\rm scatt}=(1.36\pm0.02)\,{\rm mas}\times(\lambda/{\rm cm})^2.
\end{equation}

Using a closure amplitude technique, however, Bower et
al. \cite{BowerFalckeHerrnstein2004} found that at 43 and 22 GHz the
measured sizes actually deviate slightly from the predicted
$\lambda^2$ behavior. Closure amplitudes are quantities formed by
combining the complex amplitudes in the correlated data
(``visibilities'') measured between sets of four different telescopes
such that telescope-based gain errors cancel. The closure amplitude
then provides a robust measure of the source size.  Bower et
al.~attributed the deviations from the scattering law to the
contribution of the intrinsic size, which was found to decrease with
frequency. Measurements at higher frequencies by Shen et al. and
Doeleman et al. \cite{ShenLoLiang2005,DoelemanWeintroubRogers2008a}
confirmed this trend, revealing that indeed at 230 GHz
($\lambda=1.3$~mm) the size is only 4~$R_{\rm S}$.  Fitting all data  \cite{FalckeMarkoffBower2009a} 
one finds an intrinsic size of Sgr~A*
of
\begin{equation}\label{sgrsize}
\phi_{\rm Sgr\,A*}=(0.52\pm0.03)\,{\rm mas}\times\left({\lambda/{\rm cm}}\right)^{1.3\pm0.1}.
\end{equation}
 For the parameters used here $(0.52\pm0.03)$ mas correspond to $(51\pm3)\, R_{\rm s}$ (see Figure~\ref{fig:size}, right).

\subsection{Radio variability of Sgr~A*}\label{sec:radiovar}
The radio emission itself is highly variable: rms variations of the
radio spectrum are 2.5\%, 6\%, 16\%, 17\%, and 21\% at wavelengths of
13, 3.6, 2, 1.3, and 0.7 cm respectively
\cite{FalckeMarkoffBower2009a}. Hence, variability seems to increase
in amplitude with increasing frequency. This is consistent with
adiabatic expansion of plasma blobs flowing outwards (in a jet or otherwise)
\cite{Yusef-ZadehWardleRoberts2006b,MaitraMarkoffFalcke2009}.  At the
highest frequencies
\cite{MiyazakiTsutsumiTsuboi2004,MauerhanMorrisWalter2005,MarroneBaganoffMorris2008,LuKrichbaumEckart2011a}
one sees even larger rms variations with outbursts of a factor of
several over the quiescent level.

Most interestingly, there seems to be a time lag between flares at
different frequencies: 43 GHz flares precede 22 GHz flares by about 20
min \cite{Yusef-ZadehRobertsWardle2006}. Given that the intrinsic size
differences between these frequencies is about 30 light minutes, it
would require relativistic outflows to propagate the flare from higher
to lower frequencies, if they represent density or energy enhancements
in the radio-emitting plasma \cite{FalckeMarkoffBower2009a}.  Note
that \cite{Yusef-ZadehRobertsWardle2006} interpret the lags as
highly sub-relativistic outflows, based on a further claimed NIR-radio
lag and not considering the size constraints.

\subsection{Accretion boundary conditions: radio polarization and X-rays}
\label{sec:linpol}

Aside from the mass, the accretion rate onto the black hole is the
second most important parameter, as it determines the level of
activity and can vary by many orders of magnitude.  Typically the
radiated power of a luminous black hole can be used to infer the
accretion rate, however this association becomes less linear for
radiatively inefficient accretion flows (RIAFs).  Initial estimates
for the accretion rate onto Sgr~A* ranged from
$\sim10^{-4}M_\odot/$yr \cite{CokerMelia1997} to $10^{-6}M_\odot/$yr
\cite{GenzelHollenbachTownes1994}, if Sgr~A* is accreting from the
winds of surrounding stars \cite{Ozernoy1989a}.  In the idealized case
of Bondi-Hoyle accretion \cite{BondiHoyle1944}, \revised{13}{the radial scale from
which the BH can accrete (i.e., the Bondi radius $R_{\rm B}$)} depends only
on the mass and speed of the surrounding gas, $R_{\rm B}=2GM/v^2_{\rm
  w}$.  For the stellar wind flow, early estimates of $v_{\rm W} \sim
600$ km s$^{-1}$ predicted $R_{\rm B}\simeq2.5\times10^{5}R_{\rm
  S}\simeq0.1$ pc, corresponding to $\sim 2.5^{\prime\prime}$
projected on the sky in the Galactic Center (\cite{Melia1992a}, scaled
to today's value of the mass).  From this radius inwards, the captured
material is either entirely retained or mass can be lost via a variety
of processes such as convection and/or winds
\cite{QuataertGruzinov2000a,BlandfordBegelman1999}.

The best constraints we currently have on the accretion rate at the
inner boundary near the black hole come from radio polarization
measurements, because the polarization angle changes as it passes
through a hot, magnetized medium.  The polarization vector rotates as
$\Delta \phi = {\rm RM}\times \lambda^2$, where ${\rm RM} =
8\times10^{5}\,{\rm rad} \,{\rm m}^{-2}\;\int B(s) \, n(s) \,{\rm d}s$
is the rotation measure, $B$ is the line-of-sight magnetic field, $n$
the free-electron density, and $s$ the path length along the
line-of-sight through the medium.

Bower et al.~\cite{BowerBackerZhao1999a,BowerWrightBacker1999a} found
that Sgr~A* is essentially unpolarized between 5-43 GHz, which is
quite different from what is found in quasars. They argued that the
accretion flow itself could lead to a depolarization of the radiation
by providing a very high RM, resulting in extremely fast rotation of
the polarization vector.

This idea was corroborated by the detection of circular
polarization \cite{BowerFalckeBacker1999}, which usually is much lower
than linear. Circular polarization is not affected by Faraday rotation
and created by a related process, called conversion (see
\cite{BeckertFalcke2002a} for an explanation of both processes). The observed
circular polarization has also maintained the same polarity for more than
three decades \cite{BowerFalckeSault2002a,MunozMarroneMoran2012a}.
The interpretation is that of significant Faraday depolarization by
thermal electrons in the source and a stable sense of rotation and
polarity in the magnetic field structure producing the
radiation \cite{Agol2000a,BeckertFalcke2002a,Enslin2003a,HuangShcherbakov2011a,ShcherbakovPennaMcKinney2012a}.

The detection of strong linear polarization at (sub)mm-waves soon confirmed
this picture
\cite{AitkenGreavesChrysostomou2000a,BowerWrightFalcke2003a}. Indeed,
the detected rotation measure of $RM\simeq-6\times10^{5}$ rad
m$^{-2}$ \cite{BowerFalckeWright2005a,MarroneMoranZhao2006a,MacquartBowerWright2006,MarroneMoranZhao2007}
is the highest ever found in any source. By assuming a range of
density and magnetic field profiles for accretion flows,  the
accretion rate can be constrained to lie between $\dot M\ga10^{-9} M_\odot/$yr and $\dot
M\la10^{-7} M_\odot/$yr on scales of some hundreds to thousands
$R_{\rm s}$
\cite{Agol2000a,MarroneMoranZhao2006a,MarroneMoranZhao2006a,SharmaQuataertStone2007a,ShcherbakovPennaMcKinney2012a}.

Around the time of the first polarization measurements, the
\textit{Chandra X-ray Observatory} was launched, with a spatial
resolution of $\sim0.5"$.  It discovered hot, thermal X-ray emission
associated with Sgr A* with $k_{\rm B}T=1.9$ keV, and resolved it to
be slightly extended as one might expect for the bound accretion flow
at $\sim R_{\rm B}$ \cite{BaganoffMaedaMorris2003a}. Hence, this is
likely the gas that Sgr A* is accreting from. For a sound speed
\begin{equation}
c_{\rm s}=\sqrt{{5 \over 3}{k_{\rm B}T/m_{\rm p}}}=550\,{{\rm km}/
  {\rm s}}\,\left({k_{\rm B}T /1.9\,{\rm keV}}\right)^{1/2}
\end{equation}
 one finds
again a Bondi accretion rate of 
\begin{eqnarray}
\dot M_{\rm Bondi} &= {4\pi\,n\, G^2
  M_{\rm bh}^2 c_{\rm s}^{-3}}\\ &\simeq 10^{-4}\,M_\odot\,{\rm
  yr}^{-1}\; \left({M_{\rm bh}\over4.3\times10^6 M_\odot}\right)^2
\left({n\over160\,{\rm cm}^{-3}}\right) \left({k_{\rm B}T \over1.9\,{\rm
      keV}}\right)^{-3},
\end{eqnarray}
 \revised{19}{where the density normalization
  is based on what is inferred from X-ray measurements
  \cite{BaganoffMaedaMorris2003a,EatoughFalckeKaruppusamy2013a}}.  The
discrepancy with the much lower rate nearer the black hole implied from the radio
polarization favors accretion scenarios which include mass
loss associated with the inflow (see also \cite{Wangetal2013}).

Most recently, a radio pulsar was found near
Sgr~A* \cite{EatoughFalckeKaruppusamy2013a,ShannonJohnston2013a} at a
projected distance of $0.1$ pc, which also shows a very high RM of
$7\times10^4$ rad m$^{-2}$, thereby requiring a few mG magnetic fields. This
result not only supports the idea of the Faraday screen being locally
associated with Sgr~A*, but also suggests that the accretion flow is
already starting with a high magnetization close to or even above the
equipartition value \cite{EatoughFalckeKaruppusamy2013a}.

\subsection{Accretion boundary conditions: the G2 event}
\revised{15}{ 
Finally, the discovery of an object coined ``G2'', by
Gillessen et al. \cite{GillessenGenzelFritz2012a} heading for a close encounter with
the BH, may soon provide an interesting opportunity to study basic
accretion physics. G2 was discovered as an extended object emitting
Br$\gamma$ emission in a highly elliptical orbit around Sgr A*, which
could be traced back to 2004. Recent observations
\cite{GillessenGenzelFritz2013a} have confirmed the original finding:
the object remains on track with a slightly later prediction for
pericenter passage in spring of 2014 at a nominal pericenter distance
of only 2200 Schwarzschild radii ($R_{\rm s}$). The object's compact
head shows a velocity shear of 600 km/s followed by a tail stretched
over 400 mas.}

\revised{15}{ The exact nature and origin of the object is still
  highly debated. The source could be a tenuous cloud as suggested by
  \cite{GillessenGenzelFritz2012a} or the disrupted atmosphere around
  a bound object like a star (e.g., \cite{MorrisMeyerGhez2012a}).
  Given the large shear much of the gas is already largely unbound
  even if the cloud would contain a star. Simulations
  \cite{BurkertSchartmannAlig2012a,AnninosFragileWilson2012a} predict
  that G2 will be disrupted and partially circularize at some thousand
  $R_{\rm s}$, providing an average feeding rate above
  $\sim10^{-7}M_\odot$/yr. If that is true the properties of Sgr A*
  may change soon. On the other hand, the magnitude of the effect
  depends still on many unknown parameters and the current accretion
  rate onto Sgr A*. For example: how much mass is going to accrete and
  over which time scale?  The actual free-fall time scale from
  $\sim$2000 $R_{\rm s}$ is roughly one month and the viscous time
  scale could be anywhere between months up to hundred years depending
  on the viscosity parameter $\alpha$. The latter would dilute the
  effect considerably. At the time of writing,  no discernable effect
  of the cloud has been observed.}

\subsection{Flares and variability, multi-wavelength data}

Although Sgr~A* has been known and characterized in the radio bands
for decades, it proved rather elusive in other frequencies.  Only
after the launch of \textit{Chandra} in 1999 was Sgr~A* finally
identified as a surprisingly weak X-ray source
\cite{BaganoffMaedaMorris2003a}.  During the discovery observation,
the first of what would turn out to be roughly daily X-ray flares was
also detected \cite{Baganoffetal2001}.  The $\sim$hour long timescale
for flare rise and decay provides a limit on the size scale of the
emitting region, using light-crossing arguments without relativistic
beaming, and corresponds to $\lesssim 10 R_ {\rm S}$.  A few years
later, once adaptive optics allowed for the further discovery of
Sgr~A* in the NIR, flares were also found in this band (e.g.,
\cite{Genzeletal2003,Ghezetal2004}).  In the last decade of monitoring
(see, e.g., \cite{DoGhezMorris2009},\cite{Dodds-Edenetal2011}, and
references in \cite{GenzelEisenhauerGillessen2010a}) a picture has
arisen where the steady, broadband radio and X-ray emission originates
from larger scales while the more variable submm bump and NIR
emission, as well as the X-ray flares, stem from non-thermal processes
in the plasma very close to the black hole itself.  Every X-ray flare
seems to have a simultaneous NIR counterpart, but the reverse is not
the case because smaller NIR flares may correlate with weak X-ray
flares that cannot overcome the ``blanket'' of steady, thermal X-ray
emission from larger scales.  The submm-wave emission appears to be variable on somewhat longer
timescales \cite{MarroneBaganoffMorris2008} and may not correlate as
strongly -- or at all -- with the NIR/X-ray variability, likely
because of optical depth effects. 

Until recently, the general picture developed that Sgr~A*'s spectrum
could be divided into ``quiescent'' and ``flaring'' states, based on
the apparent dichotomy of X-ray behavior and NIR flare analyses (e.g.,
\cite{Dodds-Edenetal2011}).   However, a reanalysis suggests
that the NIR flares may be consistent with a single \cite{Witzeletal2012}
power-law flux distribution, meaning that Sgr~A* is likely continually
variable down to weak flares that are indistinguishable from an
average, lowest state which one could associate with ``quiescence''.
The interpretation of the flares is an area of active debate.

The largest multi-wavelength campaign on Sgr A* was recently conducted
using \textit{Chandra} as part of the X-ray Visionary Project (XVP)
class\footnote{Sgr~A* XVP website:
  \texttt{http://www.sgra-star.com/}}.  These 35 days of observations
within one year roughly doubled the photon count compared to the prior
12 years, and tripled the number of detected X-ray flares, allowing
the first statistical tests to be performed similar to the NIR
studies.  These observations also utilized the High Energy
Transmission Gratings Spectrometer (HETGS; \cite{Canizaresetal2005})
for the first time, providing high-resolution spectroscopy in addition
to the highest spatial resolution available in the X-ray band.  As
part of this effort, an extensive network of ground and space-based
observatories were deployed, to obtain (quasi)-simultaneous data from
the radio through TeV $\gamma$-ray frequencies.  In the next section,
after introducing the theoretical context, we will briefly summarize
the latest conclusions from these campaigns.

\section{Models}\label{sec:model}
\subsection{Accretion models} 
Because Sgr~A* is the closest SMBH to us, its value as a testbed for
accretion theory has led to very active debates about the exact nature
of its emission processes and physical geometry.  Each particular
model for the emission can lead to radically different interpretations
of the data near the black hole, and impact on our ability to tease
out information about strong gravity effects, thus properly
understanding the astrophysics is a key ``foreground'' to the
gravitational scenarios.

By now it is generally accepted that \revised{9}{all massive galaxies
  contain a supermassive black hole}, but Sgr~A*'s extremely low level
of activity has provided several puzzles.  \revised{16}{First of all,
  the current accretion rate is about four orders of magnitude below
  the average accretion rate needed to grow a few million solar mass
  in a Hubble time. Secondly, the radio luminosity of Sgr A* seems to
  be well below the break in the luminosity function for radio cores
  \cite{NagarFalckeWilson2005} in low-luminosity AGN and hence marks a
  rather low state in the activity level of SMBHs in
  galaxies\footnote{\revised{20}{There have even been anthropic
    arguments on why the activity level in our Galaxy is so low
    \cite{Clarke1981a}.}}, and thirdly} the amount of gas available to
fuel the black hole would imply emission many orders of
magnitude larger than what is observed (i.e., comparing
$\sim10\%\,\dot M_{\rm Bondi}c^2=6\times10^{41}$ erg/sec to $\nu
L_\nu({350\,{\rm GHz}})\sim10^{35}$ erg/sec
\cite{HauboisDodds-EdenWeiss2012a}).

The extremely weak radio and X-ray emission ($L({{\rm 2-10\,
    keV}})\sim 2.5\times10^{33}$ erg/sec) led to a renaissance of RIAF
models, starting with the so-called advection-dominated accretion flow
(ADAF; \cite{NarayanYi1995,Narayanetal1998}) as well as others (e.g.,
\cite{Melia1992a,BlandfordBegelman1999,QuataertGruzinov2000a}).  These
models share a common theme of reducing the radiative efficiency via
either ``hiding'' the energy in less radiative protons and ions that advect it
beyond the event horizon, or in also allowing mass loss from the
accretion flow via convection and/or outflows.  Because more heat is
retained, the disk puffs up from gas pressure while at the same time
becoming optically thin and radiatively inefficient because of the low
particle densities. This low luminosity is somewhat more of a blessing
than a curse because radiation effects on the dynamics can be ignored ---
in marked contrast to the accretion disks of quasars, that require one
to include much more physics.

The original models for Sgr A* were so inefficient that they required
much higher accretion rates, but these were adapted after the new
radio polarization results became available that reduced the accretion
rate considerably (see section~\ref{sec:linpol}).  Interestingly the
one model which did not have to be changed was the one attributing the
emission to a jet (see below), which already required low accretion
rates from the innermost scales because it is in fact relatively
radiatively efficient
\cite{FalckeMannheimBiermann1993,FalckeMarkoff2000a}.

An important caveat in all these models is the unknown heating
mechanism of electrons via some form of weak coupling to the protons/ions.
This missing element of the microphysics has allowed tuning the
radiative efficiencies to a wide range of accretion rates and remains
a major source of uncertainty.  Hence, the exact value of $\dot M$ is
still model-dependent. Recent studies of the accretion flow using
sophisticated three-dimensional general relativistic hydrodynamic
 (3D GRMHD) simulations require values at the lower end of the
range to reproduce the submm-wave spectrum
\cite{MoscibrodzkaGammieDolence2009,ShcherbakovPennaMcKinney2012a,Dibietal2012,Drappeauetal2013}
from the accretion flow (but ignoring a jet contribution).  The latter
two works in particular utilize the first GRMHD code to include
radiative cooling self-consistently, and to fit the submm bump they
also favor an accretion rate of a few $10^{-9} M_\odot$/yr.

In any case, it seems that only a tiny fraction of the \revised{10}{captured material}
actually makes it down to the inner regions near the black hole, while
the rest is seemingly lost.  The newest results from the XVP campaign
provide the most direct evidence yet for such an outflow, as well as
for the source of fuel at the outer boundary.  As described in
\cite{Wangetal2013}, the diffuse emission is now resolved to be
elongated in a direction consistent with the stellar disk observed in
the NIR, strongly supporting the stellar-feeding paradigm.
Furthermore, the use of the HETGS for the first time allowed the
resolution of the Fe K$_\alpha$ complex into distinct lines.  The
measured ratio of H-like to He-like Fe lines allows the radial profile
of the gas temperature and density to be constrained.  The results
effectively rule out the ``no outflow'' scenario and further support
the various branches of ``lossy'' RIAF solutions \revised{21}{(i.e., decreasing
$\dot M$ as function of radius)}, as well as earlier
estimates of the radial profile by \cite{ShcherbakovBaganoff2010a}.
These powerful new constraints will need to be convolved into the
models of emission profiles, polarization, and eventually GR effects
as described in the following section.

\subsection{Contributions from a jet in Sgr A*}

Another major puzzle is whether Sgr~A* is launching collimated jets of
relativistic plasma, as is the norm in other known weakly accreting
black holes.  Already over 30 years ago, \cite{ReynoldsMcKee1980}
suggested an origin of the radio emission in a jet or wind (but from a
stellar-sized object).  Indeed, the properties of Sgr~A* resemble those
of flat/inverted-spectrum compact radio cores in more luminous AGN
(later also found in X-ray binaries), which can be explained as the
superposition of $\tau$=1 surfaces (the photosphere, where optical
depth becomes unity) in a stratified relativistic jet.  From
conservation laws, larger scales correspond to less dense regions \revised{23}{with
lower magnetic fields}, and thus
lower frequency emission (see, e.g., \cite{BlandfordKonigl1979}).

This issue was later revisited where the observed spectrum and
frequency-dependent size of Sgr~A* was explained as a scaled-down
quasar jet from an ``AGN on a starvation diet'', i.e., a supermassive
black hole with very low accretion rate
\cite{FalckeMannheimBiermann1993,FalckeMarkoff2000a}.  Based on the
``jet-disk symbiosis assumption'' \cite{FalckeBiermann1995}, where \revised{22}{the
total jet power simply  is}
$Q_{\rm jet}=q_{\rm j}\dot M_{\rm disk}$ and $q_{\rm j}\sim3-10\%$,
a number of concrete predictions were made
\cite{Falcke1999b,FalckeMarkoff2000a,MarkoffFalckeYuan2001}
that have consistently been borne out by observations,
\revised{1}{such as the
frequency-dependent size of Sgr~A* \cite{BowerFalckeHerrnstein2004} or
radio outbursts that travel from high to low frequencies, i.e., from
the inside-out, with relativistic speeds \cite{FalckeMarkoffBower2009}.} 

These results are also consistent with semi-analytical (e.g.,
\cite{Meier2001}) and GRMHD simulations (e.g.,
\cite{BeckwithHawleyKrolik2008a,McKinney2006}) that favor
geometrically thick accretion flows with ordered magnetic fields as
preferential launch sites for relativistic jets. \revised{-}{Indeed,
  it was recently shown \cite{MoscibrodzkaFalcke2013a} that the jets
  produced in such simulations can fully reproduce the observed
  flat-to-inverted radio spectrum when reasonable particle
  acceleration processes are considered.  Thus we finally have full
  GRMHD simulations that can reproduce inflow and outflows as well as
  the overall spectrum and size of Sgr A*.}.

However, jets have not yet been directly imaged emanating from Sgr A*
(though see \cite{LiMorrisBaganoff2013}), and although arguments can
be made about why intervening scattering can easily hide jets
\cite{MarkoffBowerFalcke2007}, alternate models have been proposed in
which the radio emission is produced in the accretion flow itself
(e.g., \cite{YuanQuataertNarayan2003}).  On the other hand, the radio
time lags discussed in section~\ref{sec:radiovar} suggest a
relativistic outflow and Sgr~A*'s non-thermal emission during bright X-ray
flares fits on the ``fundamental plane of black hole
activity''. \revised{2}{The fundamental plane connects radio- and
  X-ray emission of all types of weakly accreting black holes as a
  function of black hole mass
  \cite{FalckeKordingMarkoff2004,MerloniHeinzdiMatteo2003}. This
  includes stellar mass black holes at almost the same level of
  quiescence \cite{Markoff2005,Plotkinetal2012} as Sgr~A*.} Thus it
seems likely that Sgr~A* is underfed, radiatively inefficient, and
launching a jet (e.g., \cite{YuanMarkoffFalcke2002}).

\subsection{Flares and particle heating}

So far we have been discussing the steady emission, but the NIR/X-ray flares
have provided a third major puzzle, and an opportunity to probe
directly the plasma conditions very close to the black hole (within
tens of $R_{\rm S}$).  The fast timescale of the flares, simultaneous
NIR variability, and the flattening of the X-ray spectrum all argue
for a non-thermal process.  Most groups to date have generally focused
on magnetic processes similar to those in the solar corona, where
magnetic reconnection or stochastic processes can provide fast heating
and particle acceleration
\cite{MarkoffFalckeYuan2001,LiuPetrosianMelia2004,YuanQuataertNarayan2004},
resulting in either direct synchrotron emission in the X-ray band, or
synchrotron-self Compton (SSC) emission from the up-scattered
submm-bump photons.  However, a population of asteroids fragmented and
vaporized within Sgr~A*'s accretion flow can also instigate NIR/X-ray flares
consistent with the observations \cite{ZuvobasNayakshinMarkoff2012}.
Distinguishing between models requires strictly simultaneous data,
which is challenging to obtain.  The simultaneity of at least some
flares in NIR and X-ray, e.g., \cite{Eckartetal2004} favors the SSC
interpretation, while others favor direct synchrotron
\cite{Dodds-Edenetal2009}, and it is not unlikely that there is an
interplay between both processes.

An interesting effect of particle acceleration during
NIR/X-ray flares could be an increase in the intrinsic size of the jet
photosphere.  The size of the photosphere is strongly dependent on the
radiating particle distribution, and the predominantly thermal
distribution required by the submm bump allows for a small size even
in jet models (see, e.g., \cite{MarkoffBowerFalcke2007}).  If, however,
a bright flare results from sustained particle acceleration and that
plasma is then advected into the jet, one might be able to detect
variability in the measured size associated with NIR/X-ray flares.
Within the context of the XVP-linked multi-wavelength coordinated
campaign, VLBA observations were triggered from NIR flares, and the
first hint of variability in the radio photosphere size was observed
 \revised{-}{\cite{BowerMarkoffBrunthaler2013a} (see also \cite{LuKrichbaumEckart2011a})}. 

Another exciting highlight of the XVP campaign is the detection of
enough new X-ray flares with high enough cadence to study their
statistical properties.  These recent results are summarized in
\cite{Neilsenetal2013}, where a variability analysis of the flares 
provides clues about the innermost orbit of accreting plasma and thus
indirectly constrains the event horizon scale.  The results are still
preliminary but suggest a cutoff timescale for flares on the order of
the light-crossing time of the innermost stable circular orbit (ISCO)
and a variability component in the power density spectrum above
Poisson noise that is consistent with the same timescale.  Such
quasi-periodic oscillations (QPOs) have been predicted in some
accretion models of the flow from magnetic processes
\cite{Dolenceetal2012,ShcherbakovMcKinney2013} or Lense-Thirring
precession \cite{IngramDoneFragile2009} near the ISCO, with various
claimed detections in the past in X-ray \cite{Aschenbachetal2004} and
NIR \cite{EckartBaganoffZamaninasab2008} that have, however, not been
robust against statistical analysis \cite{DoGhezMorris2009}.  Thus
although still very controversial, the prospect of QPOs from Sgr A* is
tantalizing because they may offer another potential route (see, e.g.,
\cite{JohannsenPsaltis2011}) to study GR effects in addition to
imaging.

\section{Observable general relativistic effects}
\subsection{The shadow}

The presence of radiation so close to the event horizon and
the availability of a suitable observing technique, namely VLBI, begs
the question of what the observable effects of general relativity on the
image are?

The visual appearance of a black hole is a question that already
received attention rather early on.   For example, Bardeen \cite{Bardeen1973} calculated the shape
of a star behind a black hole. While this situation is exceedingly
unlikely and the effect is too small to be observable in practice, it
provided important clues on what to expect (see Figure~4 in
\cite{Falcke1999b} and Fig. \ref{fig:shadowcollection} for more recent
similar examples). With the emerging idea of optically thick,
geometrically thin accretion disks \cite{ShakuraSunyaev1973}
surrounding and illuminating black holes, Cunningham calculated such a
scenario with a focus on how the spectrum would be modified by
gravitational redshift and lensing \cite{Cunningham1975a}. Hollywood
\& Melia applied this framework to the infrared spectrum of Sgr~A*
\cite{HollywoodMelia1995a} for a Bondi-Holye disk and Viergutz
\cite{Viergutz1993c} calculated general images of accretion
disks. Typically these disks were truncated near the innermost
circular orbit (ISCO) of matter, and hence the main effect is the
deformation of a ring the size of the ISCO by gravitational light
bending near the black hole, rather than the effect of the event
horizon itself.

When the relevance of the submm bump in Sgr~A* was realized, Falcke,
Melia, \& Agol \cite{FalckeMeliaAgol2000,Falcke1999b}, building on the
work of Bardeen, calculated the appearance of a black hole surrounded
by an optically thin emission region on (sub-)ISCO scales, as is
likely the case in Sgr~A*. They identified what they called the
``shadow of the black hole'' as a tell-tale feature of the event
horizon superimposed over the background light, which could be
reasonably detected with (sub)mm-wave very long baseline
interferometry (VLBI) in Sgr~A*\footnote{The same term was
  independently introduced also by de~Vries \cite{de-Vries2000a} in a
  more mathematical treatment of the problem and again applied to
  optical observations of more distant black holes.}.

The shadow is essentially a lensed image of the event horizon and its
shape is closely related to the photon orbit, which describes a closed
orbit for photons around the black hole. \revised{14}{Since photons
  that circle the black hole slightly within the photon orbit will end
  up inside the event horizon while photons just outside will} escape
to infinity, there is a rather sharp boundary between bright and dark
regions. The shadow itself is thus indeed primarily a deficit of
photons due to absorption by the event horizon. Given that light rays
circling outside the photon orbit pass longer through light-emitting
material (if optically thin), there is also a possibility of an
enhanced photon ring surrounding the shadow
\cite{JohannsenPsaltis2010a,KamruddinDexter2013a}.

The diameter of the shadow is only marginally dependent on spin,
ranging from 4.5 $R_{\rm s}$ for a maximally spinning black hole to
$\sqrt{27}\,R_{\rm s}$ for a Schwarzschild black hole. This is in
marked contrast to other properties of black holes, such as the ISCO,
which strongly scales with spin. Light rays passing on the side where
they are co-rotating with the black hole come much closer to the
center of mass with increasing spin than light rays passing on the
other, i.e., counter-rotating, side. The latter have to pass at much
larger distances to avoid being caught in the event horizon. Hence,
the centroid of the shadow can shift significantly with respect to the
mass center, but shrinks only by some $\sim10\%$ as a function of
spin.

Given that the black hole location in an image is difficult to
measure, the shift itself may be hard to detect. Nonetheless, the spin
still has a marked impact on the appearance
\cite{Takahashi2004a,BroderickLoeb2006a,HiokiMaeda2009a}. For one, the
higher the spin, the closer the photon orbit and the higher the
emissivity on the co-rotating side. Secondly, the relativistic beaming
effect due to the emitting material (if co-rotating with the black
hole spin!)  will amplify the co-rotating side further and a spinning
black hole will have a much stronger asymmetry compared to a
non-rotating black hole.  The effect of spin is also dramatic in terms
of the spectral properties of the submm bump
\cite{MoscibrodzkaGammieDolence2009,Drappeauetal2013}.

\subsection{Testing theories of gravitation}

In the meantime an entire ``shadow industry'' has emerged that
proposes to test general relativistic effects using imaging
techniques.

First of all, the detection of the shadow would distinguish between
objects with and without an event horizon
\cite{BambiFreeseTakahashi2009a}. There are reasonable arguments based
on the emitted spectrum alone that an event horizon should exist in
Sgr~A* \cite{FalckeBiermann1994a,BroderickNarayan2006a,BroderickLoebNarayan2009a},
but ``seeing is believing'' in science.

Secondly, one could start to test alternatives to standard Kerr black
holes, such as super-spinning black holes \cite{BambiFreese2009a},
boson stars \cite{TorresCapozzielloLambiase2000a}, gravastars
\cite{Chirenti07}, wormholes \cite{Bambi2013a}, and massless
braneworld black holes \cite{EiroaSendra2012a}. \revised{3}{In boson
  stars, for example, the shadow is expected to be less sharp and
  super-spinning black holes have a much smaller shadow --- if any. In
  addition, of course, one always needs to verify that any alternative
  model can also address the full range of astrophysical properties
  observed in Sgr~A* and other black hole candidates. For example, if
  a specific boson mass is required, it should probably not vary
  between different galaxies.}

Moreover, it has also been suggested that non-standard GR theories
could be tested or at least constrained, such as Kerr-Taub-NUT black
holes \cite{AbdujabbarovAtamurotovKucukakca2013a}, charged
Reissner-Nordstr\"om black holes
\cite{Zakharovde-PaolisIngrosso2005a}, \revised{-}{rotating
  Ho\v{r}ava-Lifshitz black holes
  \cite{AtamurotovAbdujabbarovAhmedov2013a}}, Sen black holes
\cite{HiokiMiyamoto2008a}, Kaluza-Klein rotation dilaton black
holes\cite{AmarillaEiroa2013a}, black holes in $\delta=2$ Tomimatsu-Sato
(TS2) spacetime
\cite{BambiYoshida2010a}, and in general black holes in space times
with extra dimensions \cite{ZakharovPaolisIngrosso2012a}.  It has also
been pointed out that one can use the Sgr~A* shadow to distinguish a Kerr
hole from a naked singularity \cite{HiokiMaeda2009a} and of course
from the combination of multiple black holes
\cite{YumotoNittaChiba2012a}. The latter, however, seems very unlikely
in the Galactic Center due to the positional stability of the radio
source \cite{ReidBroderickLoeb2008a}. 


\begin{figure}
\includegraphics[width=\textwidth]{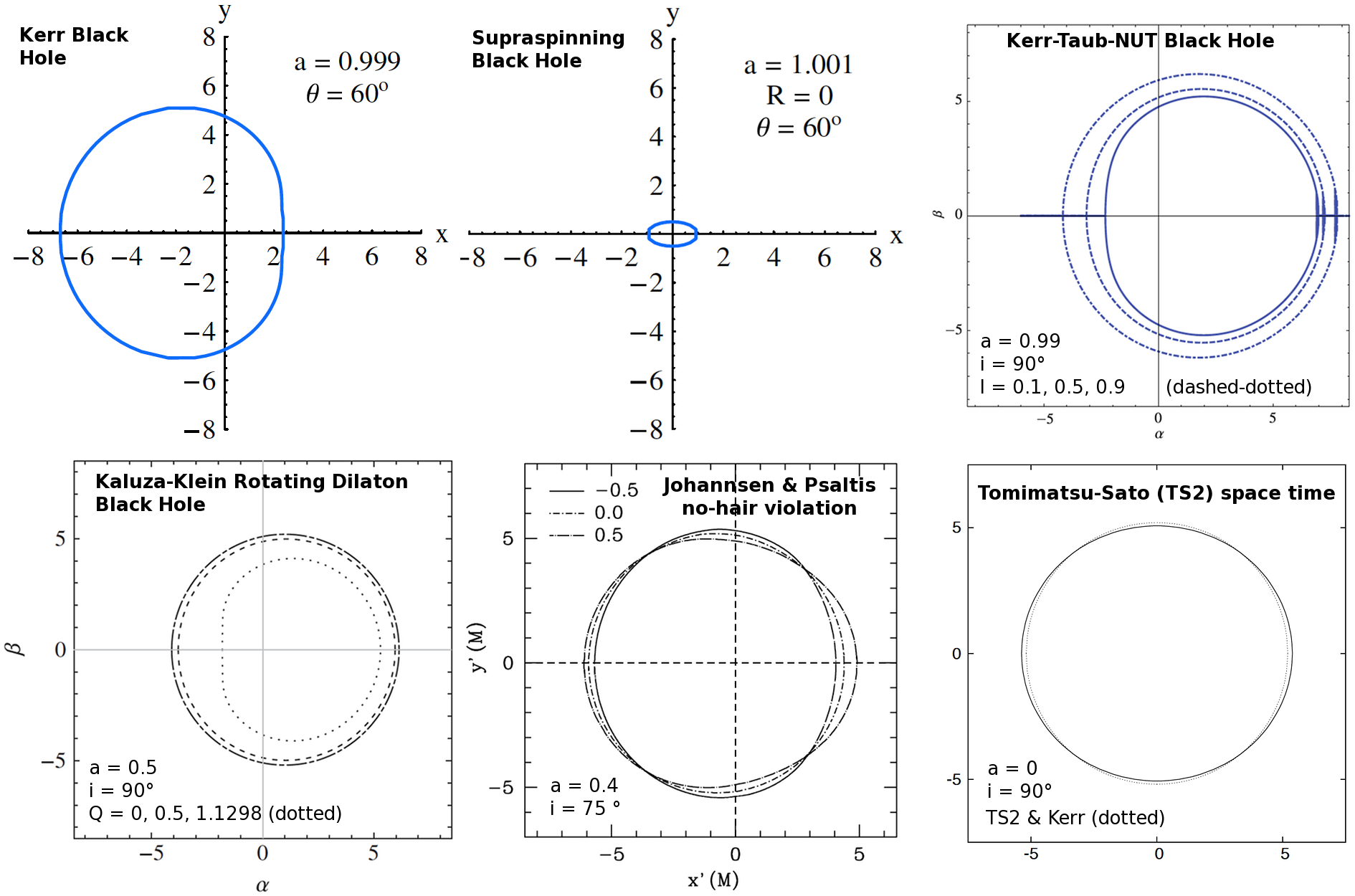}
\caption{\label{fig:shadowcollection} \revised{3}{Collection of shadow
    contours for
  different types of (non-standard) black holes. From top-left to bottom
  right: Kerr black hole (as reference),  super-spinning black hole \cite{BambiFreese2009a}, Kerr-Taub-NUT black
hole \cite{AbdujabbarovAtamurotovKucukakca2013a}, Kaluza-Klein rotation dilaton black
holes \cite{AmarillaEiroa2013a}, a parametrized no-hair violation black
hole \cite{JohannsenPsaltis2010a}, black holes in Tomimatsu-Sato spacetime
\cite{BambiYoshida2010a}. The relevant parameters as used in
the respective papers are given in the panels. If multiple lines are
plotted and multiple parameters
are listed, then the line type corresponding to the last parameter
value is also given. Note that Johannsen \& Psaltis plot the contours of the
photon ring surrounding the shadow rather than the shadow itself.
}}
\end{figure}

Johannsen \& Psaltis \cite{JohannsenPsaltis2010a} chose a slightly
different approach by allowing for a small perturbation to the
quadrupole moment (besides spin and mass) of a black hole and
calculating the images in the corresponding metric. While not born by
any theory of gravity this allows one to quantify the validity of the
no-hair theorem.
 
\revised{3}{Many of these studies are not yet quantitative enough that
  it is easy to tell how well an actual observation of Sgr~A* can
  distinguish one model from the other in the presence of instrumental
  limitations. Figure~\ref{fig:shadowcollection} shows a sample of
  different contours of black hole shadows in different
  metrics. Indeed, certain spacetimes, like Kaluza-Klein black holes,
  can produce strong changes that should be readily distinguishable,
  while others, like TS2 spacetime, only produce a few-percent wobbles
  on the shadow, which would require extremely precise measurements.
  This plethora of scenarios probably warrants a meta-study to convolve
  all these predictions with actual detector simulations, similar to
  those trying to constrain spin and inclination from current data
  based on the Kerr metric \cite{BroderickFishDoeleman2011a}. }

\subsection{The role of astrophysical models}

An important caveat is certainly the unknown structure of the
astrophysical source of emission. \revised{4}{The fact that the
  emission becomes optically thin in the submm-wave regime is well
  established by the turnover in the submm bump as well as the change
  in polarization.  In the transition region however, which will be
  observed in the first experiments at 230 GHz, optical depth effects
  can still be expected to play a role and will be somewhat model
  dependent.}  Similarly, while the existence of a shadow itself is
generic to all models, its detailed appearance and brightness
distribution will depend strongly on the underlying astrophysical
model as discussed in section~\ref{sec:model}. Hence, black hole
parameter estimates from simulations are premature. The reason why
this is a difficult issue, is that the visual appearance is not only
determined by the structure of the plasma flow alone, but also by the
assumed heating and acceleration of electrons --- a microphysical
process that is not well understood yet. One might think that
electrons are just the icing \revised{15}{on} the cake, but if the
icing is the only thing one can see, the actual shape of the cake can
quickly become a secondary factor.

Nonetheless, many shadow images have been calculated using MHD models
\cite{YuanShenHuang2006a,HuangCaiShen2007a,MoscibrodzkaGammieDolence2009,DexterAgolFragile2010a,BroderickFishDoeleman2011a,FalckeMarkoffBower2011a}
and compared to VLBI data (real or simulated) and provided reasonable
agreement. \revised{4}{The most dramatic change in expected
  emission properties is seen when taking potential emission from the jet into
  account \cite{MoscibrodzkaFalcke2013a}. However, even in this case
  the shadow structure is still discernable at 230 GHz.}  In addition,
polarization of the emission and its signatures while passing near the
black hole can provide additional information
\cite{BromleyMeliaLiu2001a,FishDoelemanBroderick2009a,HuangLiuShen2008a},
even though calculating polarized radiation transport in an ionized medium near a
black hole is far from trivial \cite{BroderickBlandford2003a}.

\subsection{The experimental side}
\begin{figure}
\includegraphics[width=\textwidth]{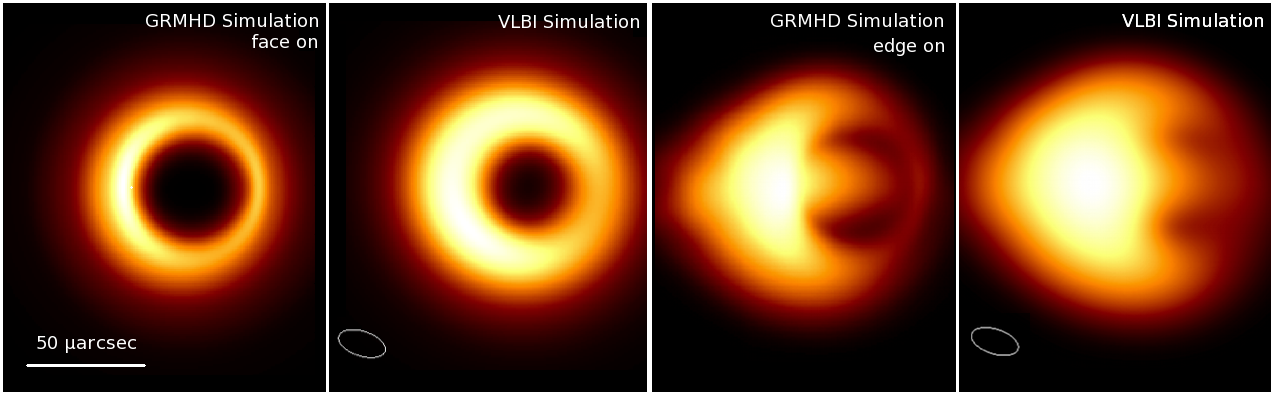}
\caption{\label{fig:shadow} GRMHD simulation
  \cite{MoscibrodzkaGammieDolence2009} of the emission in an accretion flow
  around a rapidly spinning BH in Sgr~A* blurred according to the
  expected interstellar scattering. This is compared to a
  reconstructed image from simulated submm-VLBI for face-on and
  edge-on orientations of the accretion flow
  \cite{FalckeMarkoffBower2011a}. \revised{24}{The small white ellipse
  indicates the reconstructed beams sizes.} In the optimal case \revised{25}{(face-on)}, the shadow is
  easily visible, while in the most pessimistic case \revised{25}{(edge-on)} a dynamic range
  $\ga$200:1 is needed to reveal the faint photon ring.}
\end{figure}

The shadow itself remains a rather robust feature of these simulations
unless the optical depth becomes large.  Figure~\ref{fig:shadow} shows
one example. If the orientation and spin are favorable, the shadow
should be easily visible. However, if the orientation is more edge-on
and the spin is high, the contrast between approaching and receding ring
will be high as well. 

\revised{5}{Indeed, under the assumption that the emission is
  described by a simple RIAF model, Broderick et
  al. \cite{BroderickFishDoeleman2009a} argue that face-on models are
  already highly disfavored by current data.  After all, some basic
  information about orientation and basic structure can already be
  retrieved by just combining three telescopes to measure the closure
  phase\footnote{In contrast to the closure amplitude, which is
    sensitive to the source size, the closure phase is very sensitive
    to the source structure.} at 230 GHz, which turns out to be quite constraining in ruling
  out various models \cite{BroderickFishDoeleman2011a}.  VLBI
  simulations with a larger array of (existing) telescopes have been
  published in
  \cite{DoelemanFishBroderick2009a,FishBroderickDoeleman2009a}, but
  given the systematic uncertainty in predicting the correct
  astrophysical model, there is no clear understanding yet how
  reliably black hole parameters such as spin can be determined.}
 
  \revised{5}{Clearly, three baselines will not be sufficient to
    convincingly measure a shadow. In particular, the edge-on example
    of Fig.~\ref{fig:shadow} requires much higher dynamic range in
    the imaging ($>$200:1)} and hence a sizable number of sensitive
  telescopes ($\ga6-10$) \cite{FalckeMarkoffBower2011a}.  Given that
  interferometry is very precise in measuring sizes, and the
  astrometric precision increases linearly with signal-to-noise, it
  does not seem unreasonable that deviations of $\la10\%$ from the
  expected shape could be detectable with current technology. A
  dedicated array built for just this purpose might, however, do even
  better in the future --- especially if higher frequencies ($>$350
  GHz) can be utilized, perhaps even from space.

At present one has to be content with existing submm-wave telescopes,
though major breakthroughs have already been achieved at frequencies
of $\sim230$ GHz
\cite{KrichbaumGrahamWitzel1998a,DoelemanWeintroubRogers2008a,FishDoelemanBeaudoin2011a},
where interferometry is still in its experimental phase.  Nonetheless,
around the world, groups are gearing up to combine the available
submm-wave telescopes into an ``Event Horizon Telescope'', which would
provide the highest resolution ``camera'' available in astrophysics
\cite{FishDoeleman2010a,FalckeLaingTesti2012a,KrichbaumRoyWagner2013a}.
A crucial element of this plan will be the Atacama Large Millimetre Array
(ALMA) which provides unprecedented sensitivity if properly equipped
for VLBI.

It is very important to remember that not every interferometer nor any
wavelength will do: the emission must be produced at event horizon
scales and must also be optically thin!  For Sgr~A* we know this to be
true only in the submm-wave, IR and X-ray regimes, but only submm has
the potential currently for global interferometry.  \revised{15}{Also,
  should the accretion rate increase substantially, e.g., due to
  something like the G2 event, the source could become optically thick
  even at submm-waves and, e.g., far-infrared interferometry would be
  needed}.

Apart from the Galactic Center the nearby elliptical galaxy M87, with
its powerful radio jet, may also lend itself to observing a black hole
shadow. Other black holes are too small at their respective distances
\cite{JohannsenPsaltisGillessen2012a} and would require very novel and
difficult techniques, such as X-ray interferometers \cite{Cash2005a},
which are far from likely in the current funding environment.  

\subsection{More than a shadow: stars and pulsars}

Finally, it is not imaging alone that promises significant advances in
testing general relativity in the Galactic Center. The NIR field will
advance further using interferometry together with a technique called
phase-referencing, allowing very high positional accuracy for very
faint objects.   For instance, ESO's planned GRAVITY instrument
\cite{EisenhauerPerrinBrandner2011a}  should be able to detect
stars orbiting the black hole at even tighter orbits, and provide
astrometry at the 10$\mu$as level, thereby providing even better mass
measurements.  \revised{6}{The general-relativistic periastron shift and the Lense-Thirring
precession of the orbital angular momentum will influence such stellar
orbits and, for stars passing at small distances from the BH, the
timescale of these relativistic effects is short enough to be within
the reach of GRAVITY, thus allowing one to constrain the BH spin from
stellar orbits. }

\revised{6}{Since the imaging resolution of IR interferometry using
  the VLT is on the milliarcsecond level, the shadow itself cannot be
  resolved. However, it has been suggested that the effect of orbiting
  hotspots around the black hole could lead to detectable position shifts,
  polarization variation, and position-intensity correlation that are
  expected for a source orbiting a black hole.}

Another real breakthrough would be the discovery of a pulsar orbiting
Sgr~A* on a tight orbit.  \revised{6}{In principle, it is already
  sufficient to find and time a single normal, slowly rotating pulsar
  in an orbit similar to that of stars targeted by GRAVITY, to measure
  the mass of Sgr A* with a precision of 1 solar mass, i.e., a
  relative precision of $<10^{-6}$. } Such a pulsar could in principle
also determine the black hole spin to 0.1\%, and the quadrupole moment,
testing the no-hair theorem, to 1\%
\cite{WexKopeikin1999,LiuAndOthers2012}. In general, a stellar mass
pulsar in orbit around a supermassive black hole, will probe a
completely new parameter space in relativity. Alas, until recently not
a single pulsar had been found within some ten parsecs of the Galactic
Center, despite the expectation to see thousands
\cite{WhartonChatterjeeCordes2012a}. Finding a pulsar in the Galactic
Center is perhaps one of the most difficult challenges in pulsar
astronomy, due to the strong acceleration of the source and dispersion
and scattering of the signal. Fortunately, there is some hope as the
first radio pulsar within 0.1 pc of Sgr~A* has just been found
\cite{EatoughFalckeKaruppusamy2013a,ShannonJohnston2013a}. With
sensitive telescopes such as the VLA and ALMA, the big single dishes
(Parkes, Effelsberg, GBT), and the SKA in the future, more advanced
searches may turn up many more of these.

\section{Summary and conclusion}
The good news is that there is a very strong case for a supermassive
black hole in the center of our Milky Way whose event horizon can be
imaged with radio interferometry techniques. In Table~\ref{sgrprop} we
summarize its main properties reviewed here. The mass of the black
hole is robustly determined from stellar orbits --- better than for
any other black hole candidate in the universe and confirmed by
different groups and different telescopes. The mass is also clearly
associated with a compact radio source, Sgr~A*, that is physically
located in the Galactic Center and which does not show any motion
itself --- as expected for a massive object.  The accretion rate onto
the black hole is constrained in both directions by radio polarization
observations and X-ray imaging. Sgr~A*'s low bolometric luminosity as
well as its spectrum cutting off in the submm-wave (THz) regime
supports the presence of an event horizon. Radio interferometry
measurements confirm that the radio emitting region becomes smaller as
the frequency increases and that it approaches event horizon scales at
frequencies above 230 GHz. A spectral turn-over at these frequencies
indicates that the region transitions to become optically thin and
allows us to see through to the event horizon. This effect should lead
to the appearance of a ``black hole shadow'', which is a lensed image
of photons absorbed into the event horizon. The shadow is in principle
detectable with present-day technology and would allow many
fundamental tests of general relativity and its alternatives.

Combining the information from submm-wave interferometry, broadband
multi-wavelength observations, stellar orbits, potentially the
information from pulsars, and coupled to advanced general relativistic
and magnetohydrodynamic simulations will provide us with an
outstanding astrophysical laboratory -- likely the best there is and
ever will be -- of general relativity in its strongest limit and of
black hole astrophysics.

If we ever want to know whether the dark objects lurking in the
centers of galaxies are really supermassive black holes and want to
understand how they work, the Galactic Center is the place to
study. The nearby galaxy M87 may corroborate these results in the
future, but it will always suffer from a lower confidence in
independent mass determinations and hence can only supplement studies
of Sgr~A*.  

Together with gravitational wave detections of merging black holes and
the detection of a pulsar-black hole binary, the Galactic Center
therefore promises a golden future for experimental studies of general
relativity 
\ack

HF acknowledges funding from an Advanced Grant of the European
Research Council under the European Union’s Seventh Framework
Programme (FP/2007-2013) / ERC Grant Agreement n. 227610.  SM
acknowledges support from a Netherlands Organization for Scientific
Research (NWO) Vidi Fellowship.

We thank S. Gillessen, T. Krichbaum, G. Nelemans, M. Mo{\'s}cibrodzka,
E. K\"ording, and three anonymous referees for useful comments.

\begin{table}[htb]
\caption{\label{sgrprop}Sgr~A* fact sheet}
\footnotesize\rm
\begin{tabular*}{\textwidth}{@{}l*{15}{@{\extracolsep{0pt plus12pt}}l}}
\br
Property&Value&Ref.\\
\mr
Position (J2000)&$\alpha (1996.25) = 17^{\rm h}45^{\rm m}40\fs0409, \quad\delta (1996.25) = -29\degr00\arcmin28\farcs118$&\cite{ReidReadheadVermeulen1999}\\ 
Position (Galactic)&${\rm l}(1996.25) = 359.9442085, \quad {\rm b}(1996.25) = -0.04609407$&\cite{ReidReadheadVermeulen1999}\\ 
Proper motion (equatorial) & $\dot \alpha_{\rm  east}=-3.151\pm0.05$ mas yr$^{-1 }$,  \quad $\dot \delta_{\rm  north}=-5.547 \pm 0.13$ mas yr$^{-1 }$&\cite{ReidBrunthaler2004a}\\
Proper motion (Galactic) & $\dot l_{\rm  long}=-6.379 ±\pm 0.13$ mas yr$^{-1 }$,  \quad $\dot b_{\rm  lat}=+0.202 \pm 0.02$ mas yr$^{-1 }$&\cite{ReidBrunthaler2004a}\\
Mass & $M_{\rm Sgr\,A*}=4.3(\pm0.4)\times10^6 M_\odot=8.5(\pm0.8) \times10^{36}\,{\rm kg}$&\cite{GillessenEisenhauerTrippe2009a}\\
Distance & $D_{\rm Sgr\,A*}=8.35(\pm0.15)\;{\rm  kpc}=2.58(\pm0.05)\times10^{17}\,{\rm  km}$&\cite{GillessenEisenhauerTrippe2009a}\cite{ReidMentenBrunthaler2013a}\\
Angular scales &1 mas = $1.2\times10^9\,{\rm km}$ = 98 $R_{\rm  s}$, 
\quad $1^{\prime\prime}$ = $1.2\times10^{12}\,{\rm km}$ = 0.04 pc&\\
&\quad $30^{\prime\prime}$ = $4\times10^{13}\,{\rm km}$ = 1.2 pc, \quad $10^\prime$ = 24 pc&\\
Gravitational radius &$R_{\rm g}={GM/c^2}=6.3\times10^6\;{\rm km}\sim5.1\,\mu{\rm as}$&\\
Schwarzschild radius &$R_{\rm S}={2GM/ c^2}=13\times10^6\;{\rm  km}\sim10.2\,\mu{\rm as}$&\\
ISCO for non-rot. BH &$R_{\rm ISCO}={6GM/ c^2}=3.8\times10^7\;{\rm  km}\sim30.7\,\mu{\rm as}$&\\
Orbital timescale at ISCO & $\tau_{\rm orb}\simeq2\pi Sqrt{6^3} R_{\rm 6}/c=65.2\,{\rm min} \left({M_{\rm bh}/4.3\times10^6 M_\odot}\right)$&\cite{ShibataSasaki1998a,BursaAbramowiczKaras2007a}\\
Shadow diameter (a=0) &$d_{\rm Shadow}=\sqrt{27}\, R_{\rm s}\simeq53\,\mu{\rm as}$&\\
Shadow diameter (a=1) &$d_{\rm Shadow}=9/2\, R_{\rm  s}\simeq46\,\mu{\rm as}$&\\
Measured size &$\phi_{\rm Sgr\,A*}=(0.52\pm0.03)\,{\rm mas}\times\left({\lambda/{\rm cm}}\right)^{1.3\pm0.1}$&\cite{FalckeMarkoffBower2009a}\\ 
 &$R_{\rm Sgr\,A*}=(51\pm3)\,R_{\rm s}\times\left({\lambda/{\rm cm}}\right)^{1.3\pm0.1}$&\cite{FalckeMarkoffBower2009a}\\ 
Rotation measure &$RM\simeq-6\times10^{5}$ rad m$^{-2}$&\cite{MacquartBowerWright2006,MarroneMoranZhao2007}\\
Accretion rate ($R\la100R_{\rm s}$)&$10^{-7} M_\odot/$yr $\ga\dot M\ga10^{-9} M_\odot/$yr&\cite{MarroneMoranZhao2006a,Agol2000a}\\ 
Bondi radius & $r_{\rm Bondi}={2 G M_{\rm bh}/ c_{\rm s}^2} =0.12\,{\rm  pc}\; \left({M_{\rm bh}/4.3\times10^6 M_\odot}\right)\left({c_{\rm s}/550\,{\rm km/s}}\right)^{-2}$&&\\
                    & $r_{\rm Bondi}=3\times10^5R_{\rm s} \sim 3^{\prime\prime}$&&\\
Grav. sphere of influence&  $R(M_{\rm BH}=M_{\rm stars}) \simeq 3$ pc&\cite{GillessenEisenhauerTrippe2009a}\\
Eddington luminosity & $L_{\rm edd}=4\pi G M m_{\rm p} c/\sigma_{\rm  T}=5\times10^{44} \left({M_{\rm bh}/4.3\times10^6 M_\odot}\right)$ erg/sec&\\
Radio luminosity &   $\nu L_\nu({350\,{\rm GHz}})\sim10^{35}$ erg/sec
$\sim2\times10^{-10}\,L_{\rm edd}$&\cite{HauboisDodds-EdenWeiss2012a}\\
X-ray luminosity & $L({{\rm 2-10\, keV}})\sim 2.5\times10^{33}$ erg/sec $\sim4.6\times10^{-12}\,L_{\rm edd}$&\cite{BaganoffMaedaMorris2003a}\\
\br
\end{tabular*}
\end{table}

\section*{References}
\bibliography{hfrefs,refs}
\bibliographystyle{unsrt}

\end{document}